\documentclass[%
	final,
	journal, 
    comsoc,
	letterpaper,
	oneside,
	twocolumn,
	nofonttune,%
]{IEEEtran}%
\PassOptionsToPackage{usenames,dvipsnames,svgnames,x11names,table,prologue}{xcolor}%
\PassOptionsToPackage{hyphens}{url}%
\PassOptionsToPackage{nomessages}{fp}%
\ifCLASSINFOpdf
  \usepackage[pdftex]{graphicx}
\else
  \usepackage[dvips]{graphicx}
\fi
\ifCLASSOPTIONcompsoc
   \usepackage[caption=false,font=normalsize,labelfont=sf,textfont=sf]{subfig}
\else
   \usepackage[caption=false,font=footnotesize]{subfig}
\fi

\usepackage{stfloats}
\usepackage[english]{babel}%
\usepackage[utf8]{inputenc}%
\usepackage[babel,style=english]{csquotes}%
\usepackage{hyphsubst}%
\usepackage[%
	activate={true,%
	nocompatibility},%
	final,%
	tracking=true,%
	kerning=true,%
	spacing=true,%
	factor=1100,%
	stretch=10,%
	shrink=10%
]{microtype}%
\usepackage{setspace}%
\usepackage{xcolor}%
\definecolor{todonotecol}{RGB}{250,0,0}%
\usepackage{xparse}
\usepackage[%
	colorlinks=false,%
	urlcolor=black,%
	linkcolor=black,%
	citecolor=black,%
	filecolor=black,%
	breaklinks,%
	]{hyperref}%
\usepackage{url}%
\usepackage[%
	acronym,%
	nopostdot,%
	seeautonumberlist,%
	shortcuts,%
	section=chapter,%
	toc,%
]{glossaries}%
\glsaddkey*{url}
{}
{\glsentryurl}
{\Glsentryurl}
{\glsurl}
{\Glsurl}
{\GLSurl}
\glsaddkey*{longpltwo}
{}
{\glsentrylongpltwo}
{\Glsentrylongpltwo}
{\glslongpltwo}
{\Glslongpltwo}
{\GLSlongpltwo}
\newglossaryentry{}{%
	name={},%
	plural={},%
	description={},%
}%
\newacronym{dmz}{DMZ}{demilitarized zone}
\newacronym{arp}{ARP}{adress resolution protocol}
\newacronym{it}{IT}{information technology}
\newacronym{ot}{OT}{operational technology}
\newacronym{opcua}{OPC UA}{Open Platform Communications Unified Architecture}
\newacronym{tsn}{TSN}{time-sensitive networking}
\newacronym{scada}{SCADA}{supervisory control and data acquisition}
\newacronym{vpc}{VPC}{Virtual Process Controller}
\newacronym{plc}{PLC}{programmable logic controller}
\newacronym{udp}{UDP}{User Datagram Protocol}
\newacronym{pubsub}{Pub/Sub}{Publish/Subscribe}
\newacronym{vm}{VM}{virtual machine}
\newacronym{pc}{PC}{personal computer}
\newacronym{ptp}{PTP}{Precision Time Protocol}
\newacronym{ntp}{NTP}{Network Time Protocol}
\newacronym{splc}{Soft-PLC}{Software-PLC}
\newacronym{rte}{RTE}{Real-Time Ethernet}
\newacronym{osi}{OSI}{Open Systems Interconnection}
\newacronym{mac}{MAC}{Media Access Control}
\newacronym{e2e}{E2E}{end-to-end}
\newacronym{tcp}{TCP}{Transmission Control Protocol}
\newacronym{hmi}{HMI}{Human-Machine Interface}
\newacronym{kpi}{KPI}{Key Performance Indicator}
\newacronym{mes}{MES}{Manufacturing Execution System}
\newacronym{rami4.0}{RAMI 4.0}{Reference Architectural Model Industrie 4.0}
\newacronym{tacnet4.0}{TACNET 4.0}{Tactile Internet 4.0}
\newacronym{3gpp}{3GPP}{3rd Generation Partnership Project}
\newacronym{5gppp}{5GPPP}{5G Infrastructure Public Private Partnership}
\newacronym{itu}{ITU}{International Telecommunication Union}
\newacronym{ngnm}{NGNM}{Next Generation Mobile Networks}
\newacronym{agv}{AGV}{automated guided vehicle}
\newacronym{v2v}{V2V}{vehicle-to-vehicle}
\newacronym{v2x}{V2X}{vehicle-to-infrastructure}
\newacronym{d2x}{D2X}{device-to-x}
\newacronym{qos}{QoS}{quality of service}
\newacronym{noa}{NOA}{NAMUR Open Architecture}
\newacronym{dcs}{DCS}{distributed control system}
\newacronym{m2m}{M2M}{machine-to-machine}
\newacronym{sdn}{SDN}{software-defined networking}
\newacronym{erp}{ERP}{enterprise resource planning}
\newacronym{ip}{IP}{Internet Protocol}
\newacronym{lte}{LTE}{Long Term Evolution}
\newacronym{5g}{5G}{5th generation wireless communication system}
\newacronym{harq}{HARQ}{Hybrid Automatic Repeat Request}
\newacronym{rat}{RAT}{radio access technology}
\newacronym{mc}{MC}{Multi-Connectivity}
\newacronym{fbmc}{FBMC}{Filter Bank Multicarrier}
\newacronym{gfdm}{GFDM}{Generalized Frequency Division Multiplexing}
\newacronym{sla}{SLA}{service-level agreement}
\newacronym{5qi}{5QI}{5G QoS Indicator}
\newacronym{bmbf}{BMBF}{German Federal Ministry of Education and Research}
\newacronym{nrt}{NRT}{non real-time}
\newacronym{irt}{IRT}{isochronous real-time}
\newacronym{rt}{RT}{real-time}
\newacronym{ar}{AR}{augmented reality}
\newacronym{wlan}{WLAN}{wireless local area network}
\newacronym{cpu}{CPU}{central processing unit}
\newacronym{ram}{RAM}{random-access memory}
\newacronym{ue}{UE}{user equipment}
\newacronym{ran}{RAN}{radio access network}
\newacronym{bs}{BS}{base station}
\newacronym{cn}{CN}{core network}
\newacronym{epc}{EPC}{evolved packet core}
\newacronym{mec}{MEC}{mobile edge cloud}
\newacronym{rtt}{RTT}{round-trip time}
\newacronym{vpn}{VPN}{virtual private network}
\newacronym{vpf}{VPF}{Virtual Process Control Function}
\newacronym{vdi}{VDI}{Association of German Engineers}
\newacronym{ipc}{IPC}{industrial pc}
\newacronym{rtos}{RTOS}{real-time operating system}
\newacronym{vpcmo}{VPC M\&O}{VPC Management \& Orchestration}
\newacronym{i/o}{I/O}{input /output}
\newacronym{fbd}{FBD}{Function Block Diagram}
\newacronym{ldd}{LDD}{Ladder Logic Diagram}
\newacronym{sfc}{SFC}{Sequential Function Chart}
\newacronym{il}{IL}{Instruction List}
\newacronym{st}{ST}{Structured Text}
\newacronym{os}{OS}{operating system}
\newacronym{ir}{IR}{Inactive Resource}
\newacronym{msc}{MSC}{message sequence chart}
\newacronym{icps}{ICPS}{industrial cyber-physical system}
\newacronym{cps}{CPS}{cyber-physical systems}
\newacronym{mtd}{MTD}{Moving Target Defense}
\newacronym{iot}{IoT}{Internet of Things}
\newacronym{iiot}{IIoT}{industrial Internet of Things}
\newacronym{iiot2}{IIoT}{Industrial IoT}
\newacronym{k8s}{K8s}{Kubernetes}
\newacronym{rkt}{rkt}{CoreOS Rocket}
\newacronym{dcos}{DC/OS}{Distributed Cloud Operating System}
\newacronym{ml}{ML}{machine learning}
\newacronym{ai}{AI}{artificial intelligence}
\newacronym{ict}{ICT}{industrial communication technology}
\newacronym{kvm}{KVM}{kernel-based virtual machine}
\newacronym{ie}{IE}{Industrial Ethernet}
\newacronym{ias}{IAS}{industrial automation system}
\newacronym{cots}{COTS}{commercial off-the-shelf}
\newacronym{slat}{SLAT}{Second Level Address Translation}
\newacronym{gui}{GUI}{graphical user interface}
\newacronym{l2}{L2}{layer 2}
\newacronym{l3}{L3}{layer 3}
\glsdisablehyper
\usepackage[%
	backend=biber,%
	style=ieee,%
	isbn=false,%
	hyperref=true,%
	maxbibnames=99,%
	sorting=none,%
	natbib=true,%
	language=english,%
	defernumbers=true,%
	]{biblatex}%
\DeclareFieldFormat{sentencecase}{\csname bbx@colon@search\endcsname#1}

\usepackage{balance}
\usepackage{nameref}
\usepackage{soul}
\usepackage{flushend}
\usepackage{textcomp}
\usepackage{calc}
\usepackage{xkeyval}
\usepackage{multirow}
\usepackage{tabulary}
\usepackage{makecell}
\usepackage{tikz}
\usepackage{textcomp} 
\usepackage{verbatim}
\usepackage{pgfplots}
\newcommand{\nl}{\par\noindent} 
\newcommand{\mytilde}{{\raise.17ex\hbox{$\scriptstyle\mathtt{\sim}$}}}

\newlength\textheighttemp%
\newlength\textwidthtemp%
\newlength\textheightstd%
\newlength\textwidthstd%
\newlength\textheightold%
\newlength\textwidthold%
\newlength\tempheight%
\newlength\tempwidth%
\SetProtrusion{encoding={*},family={bch},series={*},size={6,7}}
              {1={ ,750},2={ ,500},3={ ,500},4={ ,500},5={ ,500},
               6={ ,500},7={ ,600},8={ ,500},9={ ,500},0={ ,500}}
\SetExtraKerning[unit=space]
    {encoding={*}, family={bch}, series={*}, size={footnotesize,small,normalsize}}
    {\textendash={400,400}, 
     "28={ ,150}, 
     "29={150, }, 
     \textquotedblleft={ ,150}, 
     \textquotedblright={150, }} 
\SetTracking{encoding={*}, shape=sc}{40}
\usepackage{newtxmath}

\makeatletter
\let\blx@rerun@biber\relax
\makeatother
\pgfplotsset{
  grid style = {
   line width = 0.1pt
  }
}

				\newcommand{\disablewr}[1]{#1}%
				\newcommand{\newcommanddisw}[3]{\newcommand{#1}[1]{\disablewr{\textcolor{#2}{#3}}}}%
\renewcommand{\disablewr}[1]{}%
\definecolor{todocol}{named}{red}
\newcommanddisw{\todo}{todocol}{ToDo: #1}%
\definecolor{migucol}{named}{purple}%
\newcommanddisw{\migucom}{migucol}{{@}comment: #1}%
\newcommanddisw{\miguhigh}{migucol}{#1}%

\definecolor{darecol}{named}{blue}%
\newcommanddisw{\darecom}{darecol}{{@}comment: #1}%
\newcommanddisw{\darehigh}{darecol}{#1}%

	\newcommand{\TempDisplayPreparation}{\disablewr{%
		\section{Draft-State: Comment Color Code}\noindent%
		\todo{Comments: ToDos}\nl%
		\migucom{To Do and Comments: Michael Gundall}\nl%
		\darecom{To Do and Comments: Daniel Reti}

	}}%
\begin{document}%
%
\title{%
Application of Virtualization Technologies in Novel Industrial Automation: Catalyst or Show-Stopper?
\thanks{This research was supported by the German Federal Ministry of Education and Research (BMBF) within the \gls{tacnet4.0} and SCRATCh projects under grant numbers 16KIS0712K and 01IS18062E. The SCRATCh project is part of the ITEA 3 cluster of the European research program EUREKA. The responsibility for this publication lies with the authors. This is a preprint of a work accepted but not yet published at the IEEE 18th International Conference on Industrial Informatics (INDIN). Please cite as: M. Gundall, D. Reti, and H.D. Schotten: “Application of Virtualization Technologies in Novel Industrial Automation: Catalyst or Show-Stopper?”. In: 2020 18th International Conference on Industrial Informatics (INDIN), IEEE, 2020.}
}%
%
\author{%
\IEEEauthorblockN{%
    Michael Gundall\IEEEauthorrefmark{1}, %
    Daniel Reti\IEEEauthorrefmark{1}, %
    and Hans D. Schotten\IEEEauthorrefmark{1}\IEEEauthorrefmark{2} %
    \\%
}%
\IEEEauthorblockA{%
    \IEEEauthorrefmark{1}German Research Center for Artificial Intelligence GmbH (DFKI), \\ Kaiserslautern, Germany \\%
    \IEEEauthorrefmark{2}Department of Electrical and Computer Engineering, Technische Universität Kaiserslautern, \\ Kaiserslautern, Germany %
	\\%
    Email: %
        \{michael.gundall, daniel.reti, hans\_dieter.schotten%
        \}@dfki.de
        \\%
}%
}%


%

%
%
%
%
%
%
%
%
\maketitle
%
\begin{abstract}%
Industry 4.0 describes an adaptive and changeable production, where its factory cells have to be reconfigured at very short intervals, e.g. after each workpiece. Furthermore, this scenario cannot be realized with traditional devices, such as programmable logic controllers. Here the use of well-proven technologies of the information technology are conquering the production hall (IT-OT convergence). Therefore, both virtualization and novel communication technologies are being introduced in the field of industrial automation. In addition, these technologies are seen as the key for facilitating various emerging use cases. However, it is not yet clear whether each of the dedicated hardware and software components, which have been developed for specific control tasks and have performed well over decades, can be upgraded without major adjustments. 

In this paper, we examine the opportunities and challenges of hardware and operating system-level virtualization based on the stringent requirements imposed by industrial applications. For that purpose, benchmarks for different virtualization technologies are set by determining their computational and networking overhead, configuration effort, accessibility, scalability, and security.  

\end{abstract}%
%
%
%
%
%
\IEEEpeerreviewmaketitle
%
%
%
%
%
%
%
%

\section{Introduction}%
\label{sec:Introduction}
\miguhigh{Industry 4.0 is a vision that has been launched by a German initiative in 2011, and describes both the use cases and the corresponding goals of a smart manufacturing, that are time, resource and energy minimization, optimization of the product quality, and lot size one \cite{b1}.} The \gls{iiot}, in conjunction with \glspl{icps} are seen as the basis for the realization of a smart manufacturing, twhich introduces novel use cases that bring new challenges for both communication and automation systems. \miguhigh{In addition, every goal and individual use case brings up new challenges for communication and automation systems.} These can be mobility requirements, the application of computing-intensive algorithms, such as for \gls{ml} or \gls{ai}, and a highly flexible reconfiguration of manufacturing systems or entire factories. Here, a reconfiguration or even redeployment of industrial automation systems at very short intervals, e.g. before each workpiece, is conceivable, while recent industrial installations are typically based on dedicated hardware controllers, such as \glspl{plc}, that cannot provide these capabilities \cite{6246692}. In order to address these challenges, virtualization is a suitable approach. The virtualization of industrial automation systems enables the application of concepts that are already well-known in the field of \gls{it}. Since most of these concepts cannot be applied “out of the box", the question arises how these concepts can be adopted for industrial purposes.

Furthermore, the convergence  between  \gls{it} and \gls{ot}, and the connection to the \gls{iiot} makes security an important topic, which until now has not had much relevance in the \gls{ict}\cite{8401919}. This was not intended as these systems were inherently secure against network intrusion, as there were no connections to the Internet and only few wireless connections. As security mechanisms should not compromise the performance of the system, in many cases existing solutions cannot be used. Therefore, security has to be considered already in an early phase of new installations and has been a concern in our investigations. In this paper, the following contributions can be found:
\textbf{
\begin{itemize}
    \item Identification of challenges and opportunities imposed by industrial use cases and the most important requirements.
    \item Evaluation of different virtualization technologies and configurations based on the identified requirements.
\end{itemize}
}

Therefore, the paper is structured as follows: 
Sec. \ref{sec:Background} motivates our work on this topic, while Sec. \ref{sec:Industrial Automation} provides insights into challenges in industrial automation. An overview of virtualization technologies is given in Sec. \ref{sec:Virtualization Concepts}, followed by a benchmarking of the proposed 
technologies and specific configurations (Sec. \ref{sec:Virtualization Strategy for Virtualized Process Controllers}). 
Finally, a conclusion of the paper is given in Sec. \ref{sec:Conclusion}.

\section{Background}%
\label{sec:Background}
In order to allow a smart manufacturing, a multitude of novel use cases have to be realized. Due to the mobility requirements of devices and process equipment, the number of wireless connections between devices in the industrial landscape will increase rapidly. Therefore, the authors in \cite{8502649} describe a variety of emerging use cases that rely on wireless communications. Here, it must be taken into account that these devices often only have limited resources available. For instance, if \gls{ml}, \gls{ai}, or advanced control algorithms \miguhigh{such as model predictive control} shall be applied, the computing power of these resource constrained devices is not sufficient. To overcome this issue, the application of computation offloading is a suitable approach. In this concept, a cloud service performs the calculation of the complex algorithm and only transfers the output values to the resource constrained devices \cite{Mao2017MobileEC}. Another point is the limited battery life time of mobile applications, such as mobile robots or drones for industrial inspection. If the complexity of the algorithm to be performed exceeds a certain level, it is beneficial not to execute it on the mobile device itself, but to offload it to a cloud instance \cite{8254628}. Therefore a longer battery life time of the device and an increase of its availability and productivity can be achieved. 

Another area where the application of virtualization technologies is advantageous is the virtualization of industrial automation systems. Here, a number of benefits can be achieved by applying these approaches. One of these is the access to high performance computer systems that enable the previously mentioned gains of computing offloading. On the other hand, the virtualization of devices and applications offers a further level of abstraction with which both their flexibility and their availability can be enhanced.  In order to evaluate this, \cite{6837587} have evaluated a virtualized \gls{plc} based on \glspl{vm} that are running on Windows. By using this setup a variety of applications can already be realized. However, they limited the range to applications that do not have stringent requirements on latency and determinism, the so-called soft \gls{rt} applications. In this context, the introduction of container-based virtualization can offer benefits. For this reason, \cite{goldschmidt2018container} proposes a container-based architecture for flexible industrial control applications and provide performance benchmarks for containers running on a Raspberry Pi 2 with and without additional load. However, they assume an interval of 1ms for the cyclic execution of an application, which may be too high for some industrial applications. Therefore we vary the interval within the limits of \mbox{1ms - 1µs}. In addition, \cite{10.1145/2851613.2851737} investigated how the determinism of applications running in and outside a container is related to a given \gls{rt} priority and measured the handling latency of the virtual network interface of containers for the standard network drivers, thus determining the overhead of virtualization on the network interface. As it can be advantageous for industrial applications to use a network driver other than the standard one, we have included all of them in our investigations. 

\section{Industrial Automation}%
\label{sec:Industrial Automation}
Very characteristic for industrial applications are the high requirements on the cycle time, determinism, and availability, that vary tremendously from applications on the office floor. In this area, closed loop motion control is the most demanding use case group \cite{8502649}. This includes use cases such as machine tools, packaging, and printing machines, that typically require a maximum cycle time of \mbox{0.5 - 2ms}, a synchronicity of \mbox{1 - 5µs} and a maximum failure of less than one minute per year.

Therefore, the IEC~61131-3 describes the functionality and programming of \glspl{plc}. \glspl{plc} are sophisticated hardware controllers, that usually provide low processing times and a high availability. In addition to the processing time of the \gls{plc}, the cycle time also includes the duration of the data transmission from the sensors to the \gls{plc} and the reception of the output values by the actuator. Consequently, the underlying communication system also has to fulfil these demands.
In order to reliably satisfy the requirements of all use case types, very heterogeneous communication protocols, the so-called \gls{ie} protocols, were developed. These protocols are typically divided into the following \gls{rt} classes \cite{7883994}:
\begin{itemize}
\item \gls{rt} class A: $t$\textsubscript{cycle}\textless 100 ms,
\item \gls{rt} class B: $t$\textsubscript{cycle}\textless 10 ms, and
\item \gls{rt} class C: $t$\textsubscript{cycle}\textless 1 ms.
\end{itemize}
In particular, protocols targeting \gls{rt} class C are often not based on the \gls{ip} layer, which represents the network layer  (\gls{l3}) of the \gls{osi} model, but build directly on top of the \gls{mac} layer \cite{7883994}. This corresponds to the data link layer (\gls{l2}) of the \gls{osi} model. This method saves the overhead of \acrshort{udp}/\acrshort{ip} or \acrshort{tcp}/\acrshort{ip} header, which has several advantages. Firstly, more user data can be transmitted within the same packet size or the packet size can be reduced. Consequently, the number of devices can be increased and the data packets can be processed faster. For this reason, novel communication technologies and communication protocols such as \gls{tsn} 
and the \gls{opcua} standard 
with its \gls{l2}-based \acrlong{pubsub} pattern for \gls{rt} traffic are also based on these benefits \cite{opcuart}. The possibility of \gls{l2} communication is therefore a prerequisite if use cases of \gls{rt} class C are to be addressed with the virtualized industrial automation system.

In  the  case  of  mobile  use  cases, high  demands  arise  on wireless communications that cannot be met by state of the art solutions. Here, \gls{5g} is seen as a highly promising candidate. In order to integrate \gls{5g} in the industrial landscape, \cite{8731776} proposed a concept that foresees to present the \gls{5g}  system to the \gls{tsn} system like any other \gls{tsn}-aware bridge.


 \begin{figure*}[htbp]
\centerline{\includegraphics[width=\textwidth]{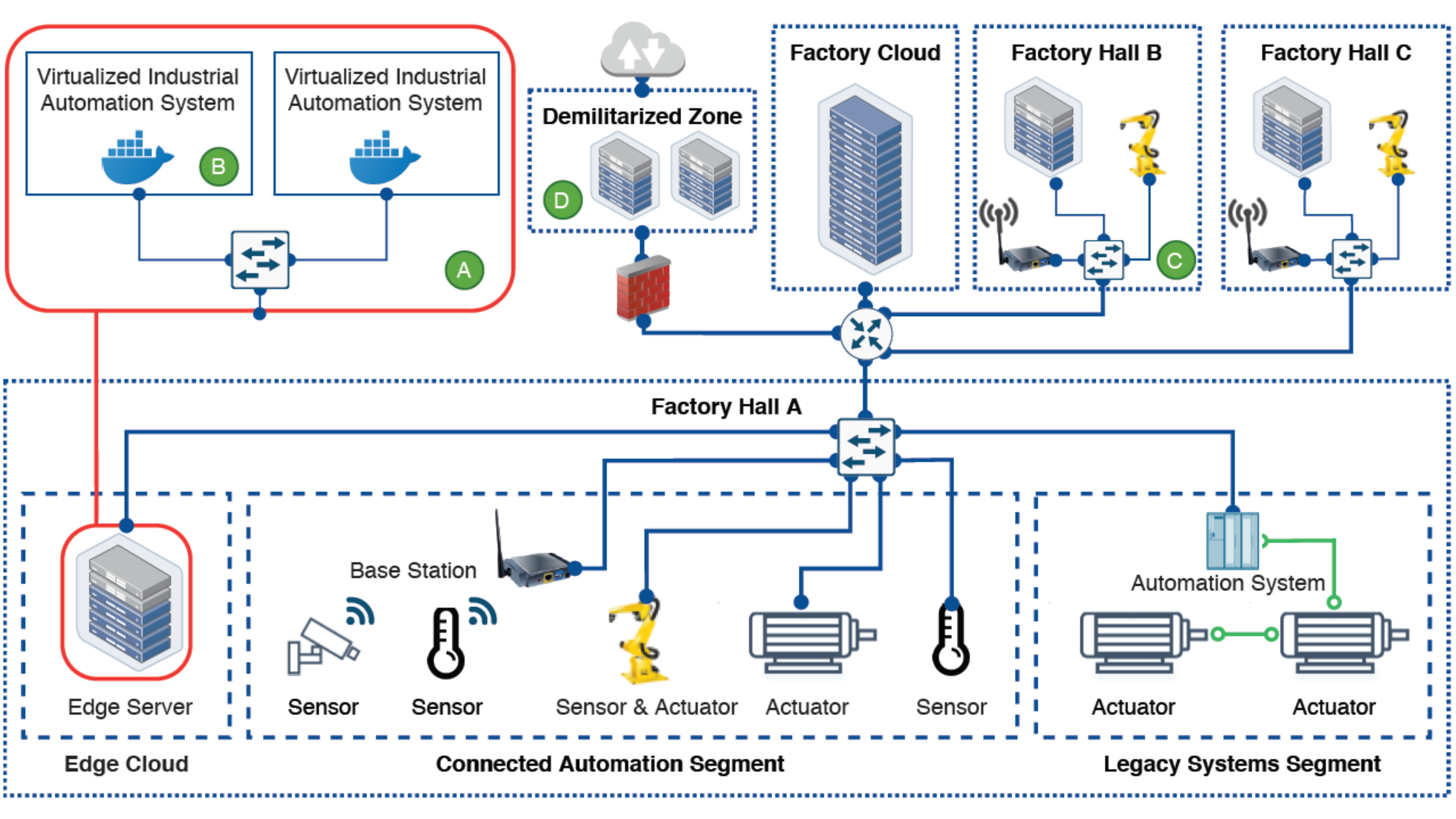}}
	\caption{Industrial automation network including possible attacker locations on a generalized network threat model.}
\label{fig:threat}
\end{figure*}
With the emerging interconnectivity of \gls{ot}, the attack surface is growing as well, requiring security considerations to be part of the infrastructure design process \cite{8502649}.
In order to evaluate the security of different network configurations a threat model for a generalized industrial automation network topology was made (see Fig. \ref{fig:threat}). The network consists of three factory halls, a factory cloud, and a \gls{dmz} that are connected over a router. In the \gls{dmz} servers are reachable from the internet and only have a limited connection to the internal network. In each factory hall an edge server is located where multiple 
virtualized industrial automation systems run on. Within a factory hall all devices are connected over a L2 bridge.  Four types of attackers are defined depending on their location in the network. As shown in Fig. \ref{fig:threat} the weakest attacker, here defined as D, can only reach the \gls{dmz} of the network from the internet, attacker C is located in the factory network but on a different subnet. This attacker can enumerate the network and try to pivot to different hosts. Attacker B has compromised a virtualized industrial automation system on the edge server and aims, for example, to escape the container restrictions and to laterally move to other hosts. Attacker A has already compromised the edge server and can, with the respective privileges, control the running container daemon and therefore change the containers configuration and connect to them. From here the attacker could steal information, cause outages (denial of service) or manipulate sensor and actuator data \cite{Zhu.2011} to damage the machinery and thereby pose fatal risk to the factory staff.

\section{Virtualization Concepts}%
\label{sec:Virtualization Concepts}

 \begin{figure}[htbp]
\centerline{\includegraphics[width=\columnwidth]{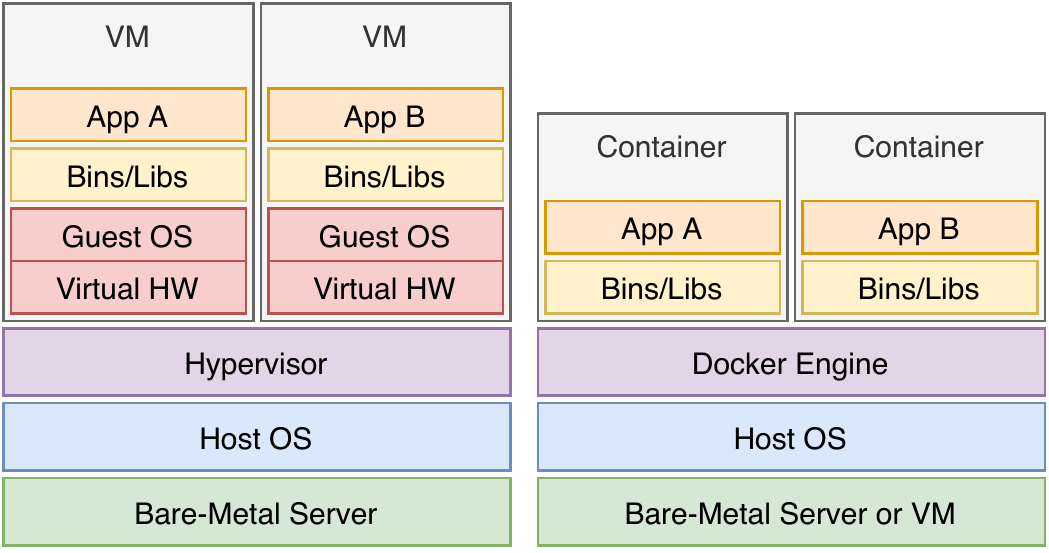}}
	\caption{Comparison of abstraction layers between \glspl{vm} and containers.}
\label{fig:VM}
\end{figure}

Virtualization refers to the replacement of physical resources with virtual replicas. In enterprise environments the main virtualization technology is server virtualization, where traditionally a software called hypervisor may run multiple virtual guest machines on the same physical hardware of the host machine. In the recent decade \gls{os}-level virtualization gained momentum in the \gls{it}, where only part of the \gls{os} is virtualized, allowing more efficiency and lower latency through a reduced overhead \cite{7095802}. Virtualization offers numerous advantages, starting with the ease of deployment, the more economically efficient use of physical hardware, and the improved security enabled by the isolation which comes along with virtualization. The following will give a brief introduction to the technologies of \glspl{vm} and containers and discuss their respective implications. Note that further, less common virtualization technologies, e.g. micro-virtualization \cite{microvirtualization}, exist but are not viable for the virtualization in the industrial context.

\subsection{\Acrfull{vm}}
Hardware virtualization software, namely hypervisor or virtual machine monitor (VMM) have first been popular in the 1970ies, when it was allowing multiple tenants to share expensive hardware. The hypervisor may run as an application on top of a running host \gls{os} or run on bare-metal with a custom kernel providing a higher performance. On top of the hypervisor multiple guest \glspl{vm} can run in parallel, whereby each \gls{vm} can have a different \gls{os} and virtual hardware. The hypervisor translates virtual \gls{cpu} instructions to the underlying hardware. To enforce the  \gls{vm} isolation, privileged instructions are intercepted and their effect is emulated on the virtual hardware. This instruction set virtualization can be achieved completely in software where the complete instruction stream is just-in-time translated, or in hardware, in case the \gls{cpu} supports virtualization, where the \gls{cpu} may take over the execution of the \gls{vm} instructions, until a privileged instruction is tried to be executed, which is then handed over to the hypervisor \cite{uhlig2005intel}. Similarly, memory virtualization can be achieved either in software or in hardware. Modern \glspl{os} already isolate processes by default through assigning virtual memory. \miguhigh{In order to avoid the steps of translating the guest’s virtual address to the guest’s physical address before translating it to the hosts address on every access, the hypervisor traces the guest’s physical address access and maps it directly to the host machine pages through a shadow page table. The overhead of cache misses through context switches is avoided in hardware through \gls{slat}, allowing to look addresses up in hardware automatically.}

\subsection{Container}
The term container refers to a virtualized environment on OS-level which offers similar advantages to \glspl{vm} regarding isolation and ease of deployment while costing significantly less resources, as multiple isolated containers can run directly on the underlying OS and physical or virtual hardware simultaneously with other containers \cite{8399557}. 
As can be seen in Fig. \ref{fig:VM}, an application and its dependencies can be packed into a container, which is running on a container management software. The user-interface for containers is provided by the container manager, offering functionality for building, managing, sharing and running containers \cite{Combe.2016}. The system independence and easy deployment along with container orchestration software, such as Docker Swarm or \gls{k8s}, allows to automate the management, deployment and scaling of containerized applications and is therefore popular for running applications in a cloud-computing infrastructure \cite{8399557}.

\miguhigh{Container isolation is achieved through two UNIX kernel features called kernel namespaces and control groups (cgroups). This isolation ensures, if the container is configured correctly, that no other processes may access the containers resources. Namespaces allow for processes to have their own local instances of kernel internal objects and paths of the file system. The cgroups feature allows to control and restrict the access to system resources for groups of processes. The restricted system resources might be \gls{cpu} time, memory access, networking capabilities, disk I/O, etc. Isolation of the file system is solved with a Union Filesystem, where the image layers of the filesystem are read-only and an overlayed layer  stores the changes made. This way multiple containers can share the underlaying image as it remains unchanged which saves storage and allows a faster deployment.}

In summary, containers offer an efficient and lightweight virtualization and allow an easy and scalable platform independent deployment of server applications and bring up many novel automation tools assisting the operation, but introduce complexity and overhead compared to bare-metal, allow only Linux as a guest system, and lack support for a \gls{gui} which some applications may require. 

\subsection{Summary}
To conclude, containers can compete with \glspl{vm} by providing similar features with better performance. Still \glspl{vm} will not disappear as they provide functionalities containers can not such as hosting guests with different \gls{os} kernels, offering a better isolation, in spite of which \gls{vm} escapes existing, and supporting applications with \glspl{gui}. While these functionalities are not relevant for most cloud services, the current practice tend towards containers running on virtualized infrastructure, inside of \glspl{vm}. In the long-term containers running on bare-metal are completely capable of replacing \glspl{vm} in cloud infrastructure, also from a security and privacy perspective. Therefore, a virtualization of industrial automation systems based on container technology is well suited for industrial applications. Boettiger motivates the usage of Docker to enable reproducible and re-usable research \cite{10.1145/2723872.2723882}.

\section{Virtualization in Future Industrial Automation Systems}%
\label{sec:Virtualization Strategy for Virtualized Process Controllers}
This section investigates how Docker containers can be used for industrial applications and compares performance benchmarks for specific configurations. First, the computational overhead associated with the use of containers is determined (Sec. \ref{subsec:computing performance}). Therefore, we compare the determinism of an application running identically in a container, in a \gls{vm}, and on a bare-metal server, based on different execution intervals.
In addition, Sec. \ref{subsec:networking performance} evaluates the different networking configurations of a Docker container, and compares several characteristics with \glspl{vm} and bare-metal. 
For all investigations we have used the equipment listed in Tab. \ref{tab:hardware}. 

\begin{table}[htbp]
\caption{Hardware configurations}
\begin{center}
\begin{tabular*}{\columnwidth}{|c|c|p{0.6\columnwidth}|}
\cline{1-3} 
\textbf{\textit{Equipment}} & \textbf{\textit{QTY}} & \textbf{\textit{Specification}}\\
\cline{1-3} 
Mini PC & 2 & Intel Core i7-8809G, 2x16 GB DDR4, Intel i210-AT \& i219-LM Gibgabit NICs, Ubuntu 18.04 LTS 64-bit, \linebreak Linux 4.19.103-rt42  \\
\cline{1-3} 
Switch & 1 & 8-Port Gigabit Ethernet Switch\\
\cline{1-3} 
\end{tabular*}
\label{tab:hardware}
\end{center}
\end{table}

\subsection{Computation Performance}
\label{subsec:computing performance}
This section examines the use of a general purpose \miguhigh{bare-metal} server for industrial applications. Here, latency and determinism are the main candidates to be analyzed. 
Therefore, we use \textit{Cyclictest}\footnote{For further information and download see the following website: https://wiki.linuxfoundation.org/realtime/documentation/howto/tools/cyclictest/}, a well-known micro-benchmarking software for the verification of \gls{rt} characteristics for x86 and x64-based systems and is also the basis of the investigations in \cite{10.1145/2851613.2851737, goldschmidt2018container}. It determines the jitter of a periodically executed task. In order to determine the performance overhead of container and \gls{vm} virtualization compared to bare-metal 
the Cyclictest programm is executed on one of the \miguhigh{\gls{cots}} mini PCs listed in Tab. \ref{tab:hardware}. We decided to use mini PCs because they are comparable to typical industrial \glspl{plc} (e.g. S7-300 \cite{s7300} and its successor S7-1500 \cite{s71500}) in terms of space consumption and pricing. Therefore, we ran the following command, where \texttt{-N} prints the result in nanoseconds, 
\texttt{-n} uses \texttt{clock\_nanosleep}, \texttt{-p} sets the \gls{rt} priority, \texttt{-D} denotes the duration of the test, and \texttt{-i} sets the interval of the execution:
\texttt{cyclictest -N -n -p99 -D10m -iI}.
Since the dependency of Cyclictest on the basis of the priorities was already part of the investigations in \cite{10.1145/2851613.2851737}, we have set the priority to 99, representing the highest level. In addition, the behavior for different intervals should be investigated, corresponding to the cyclic execution of industrial applications. Based on the requirements of the use cases in Sec. \ref{sec:Industrial Automation} we will test the behavior for intervals of 1ms, 100µs, 10µs, and 1µs. The duration of each test run is set to 10 minutes. In addition, each test run was performed for the following setups:
\begin{itemize}
    \item \textbf{No preemption:} in this configuration, the Cyclictest software module runs directly on the generic Linux kernel that was supplied with Ubtunu 18.04 LTS.
    \item \textbf{Preemption:} in this mode, a \gls{rt} patch for the Linux kernel 4.19. was applied  \cite{yodaiken1997real} and the “Fully Preemptible" mode has been activated \cite{dietrich2005evolution}.
    \item \textbf{Container:} in this setup, Cyclictest runs inside a Docker container that has been launched with the \texttt{--privileged=true} flag.
    \item \textbf{Virtual machine:} the host PC maintains a \gls{kvm}, that is also running on Ubuntu 18.04 LTS in combination with the above mentioned Linux kernel in combination with the \gls{rt} patch.
\end{itemize}

\begin{figure}[htbp]
\resizebox{\columnwidth}{!}{%
\input{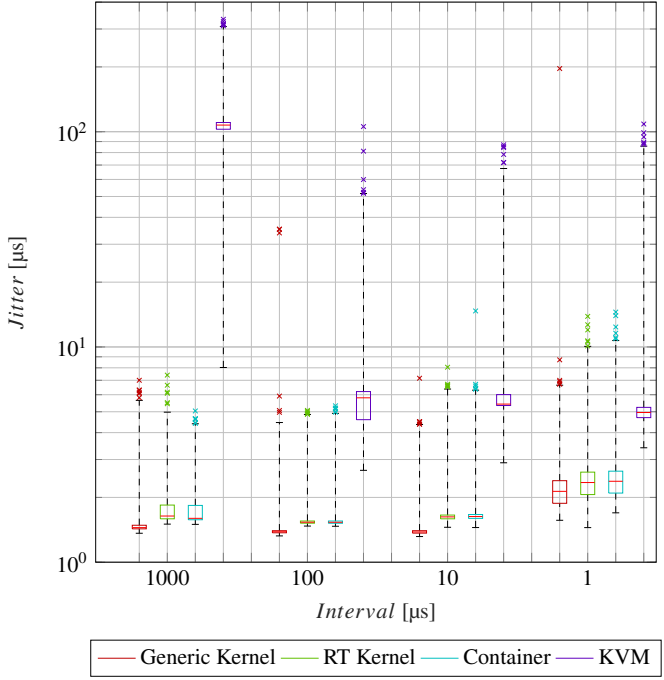}
}
	\caption{Readings for the Cyclictest software module running on different platforms and and at different intervals}
\label{fig:Cyclictest}
\end{figure}

For the graphical interpretation of the results, box plots containing 60,000 samples per data series have been created (see Fig. \ref{fig:Cyclictest}).

By default, 50\% of the values are located inside the boxes and the outliers are those values that exceed $\pm 1.5 \times IQR$. Assuming that every second value is close to the median value and the next value is significantly higher, this could lead to the worst case scenario where no values are placed between the boxes and the outliers and therefore only 50\% of the samples lie within the box plots. Since this or similar cases are not sufficient for the rigorous demands of industrial applications, we have selected the outliers so that 99.99\% of all data points are located within the box plots.

It turns out that the timing of the \gls{vm} is significantly worse than the timing of the other systems. Especially the interval of 1ms shows that the median value is 10 times higher compared to the others. The median value decreases with higher processor load, but there are still many outliers in the range of \mbox{10 - 100µs}. This confirms the previous investigations, which have shown that \glspl{vm} are not suitable for low-latency \gls{rt} applications due to the overhead by the additional software stack \cite{7095802,7164727}.

Looking at the system with the generic kernel, it can be seen that the median value of this system is the lowest, i.e. the majority of the recorded data values is below that of the other systems. However, if the processor load is increased by reducing the interval, there are some outliers upwards without the use of preemption. This corresponds to the assumption that generic kernels are generally more efficient, but cannot guarantee the level of determinism expected by industrial applications \cite{1137388}.

Next, the \gls{rt} kernel running on the bare-metal server and the deployment of a container should be compared. Here, the investigations show that there are no significant limitations when using Docker for \gls{os}-level virtualization. In an interval of 1ms the readings are even slightly lower when running the container versus using preemption, but the reason for this is probably the number of samples. Looking at the other intervals, there is a performance difference of \mbox{\flq 0.05µs} in the median values and \mbox{\flq 1µs} for the maximum values. This means, there is no significant performance impact by using a Docker container. These results comply with the findings in \cite{10.1145/2851613.2851737,goldschmidt2018container}.

Another point to consider is the required determinism. If, for example, a reliability of 99.99\%, i.e. a failure of 0.01\% of all values, is sufficient for an application, the use of a generic kernel can be advantageous as it brings a higher overall efficiency. In this case, a generic kernel can achieve a jitter of \mbox{\flq 7µs} with a reliability of 99.99\%, while an \gls{rt} kernel is only able to provide about \mbox{\flq 10µs}.

\subsection{Networking}
\label{subsec:networking performance}

The next step is to check the network performance of Docker containers for different configurations of the network interface. Furthermore, we assume the following two scenarios:
\begin{enumerate}
    \item Host-only communication
    \item Inter-host communication 
\end{enumerate}
In the first scenario, the host-only communication, there are two applications on a single host server that interact with each other. This scenario is useful if an industrial automation system needs to be reconfigured. If, for example, an update of the configuration, firmware or software is required, it is preferable to have both applications on the same system so that the migration between the old and the new application can be triggered without delay by an additional communication network. In cases, where a resilient operation of industrial automation systems is required, however, a redundant instance of the industrial automation system should be placed on a separate host in order to maintain operation of the production line in case of breakdown of the host server. This situation requires inter-host communication.

To enable one of the major advantages of virtualization technologies an automated deployment of industrial automation systems on multiple hosts should be possible. Various orchestration tools are available for a massive and automated container deployment. In this context, \gls{k8s} and Docker Swarm are the most prominent candidates. Since \gls{k8s} works on the basis of proxy servers on \gls{ip} level and a plain \gls{l2} communication is not envisaged \cite{marmol2015networking}, the investigations have been done on the basis of Docker Swarm. 
Various settings can be made when configuring Docker containers and using Docker Swarm. Of particular interest here are the network drivers. These vary depending on configuration effort, accessibility, scalability, security level, and performance. All supported network drivers are listed below:

\begin{itemize}
    \item \textbf{None:} for this container, networking is disabled. Communication outside the container is not possible. Therefore this network driver is not considered further.
    \item \textbf{Host:} in this configuration all network isolation is deactivated. All containers have direct access to the interfaces of the host and can communicate with each other as well as with external devices.
    \item \textbf{Bridge:} the default network driver. If no driver is specified when starting a container, it is assigned to this network type. All containers on the host are connected by a network bridge and can communicate. However, inter-host communication is not foreseen by the default bridge network. This driver is mainly used for standalone containers.
    \item \textbf{Macvlan:} macvlan networks allow the assignment of \gls{mac} addresses to containers so that they appear as physical devices on the network. The Docker daemon forwards the traffic to containers based on their \gls{mac} addresses. This allows containers that use the macvlan driver to avoid the routing through the network stack of the Docker host.
    \item \textbf{Ipvlan:} ipvlan can be compared with the network driver macvlan and offers a \gls{l2} and \gls{l3} mode. In contrast to the network driver macvlan, the endpoints using ipvlan have the same \gls{mac} address. 
    \item \textbf{Overlay:} overlay networks connect the Docker daemons of multiple hosts and enable the container-to-container communication of multiple hosts. By using overlay networks, the routing of packets is handled by Docker and makes routing at \gls{os}-level obsolete. For this reason this is the default network driver for Docker Swarm services.
\end{itemize}
As already described in Sec. \ref{sec:Industrial Automation}, the requirements for industrial use cases are usually expressed by the cycle time, which consists of the time to exchange the messages and the computing time of the algorithm. Since this time includes both the reception of the sensor values and the transmission of the control outputs to the actuators, the \gls{rtt} for the respective network drivers is determined and compared with bare-metal and \glspl{vm}. Once more, box plots with 60,000 data samples per series were generated for the evaluation of these measurements. Again, the outliers were defined in such a way that 99.99\% of all values are within the outer limits of the box plot.

\begin{figure}[htbp]
\resizebox{\columnwidth}{!}{%
\input{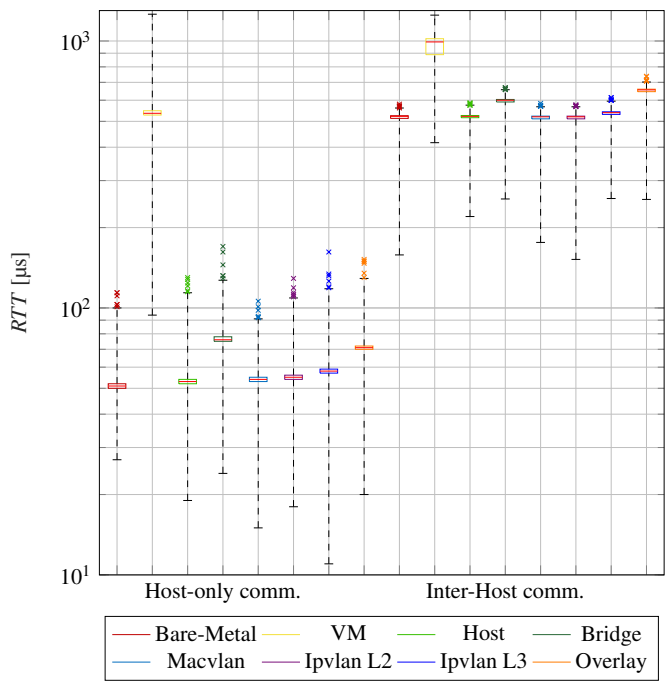}
}
	\caption{Readings for the measurement of the RTT for for host-only and inter-host communication}
\label{fig:Pings}
\end{figure}
As shown in Fig. \ref{fig:Pings}, \glspl{vm} perform significantly worse than the other configurations in both host-only and inter-host tests. In host-only communication, the median value for the RTT is about 10 times higher than in all other network configurations. In addition, the median values for the host, macvlan and ipvlan configuration differ from the bare-metal system by only 2µs and 3µs for the host-only measurements and are equal for inter-host communication. The maximum values for these network configurations also only vary in the range of 582µs $\pm$ 5µs. This means that selecting these network drivers for the deployment of a Docker container will not degrade network performance compared to the use of bare-metal.

Next, looking at the standard drivers for standalone containers and Docker Swarm clusters using bridge and overlay modes, it can be seen that these are about 20\% slower for both of the investigated scenarios. From a performance point of view, the network drivers host, macvlan and ipvlan are preferable, as they have a very low overhead and therefore no significant effect regarding the RTT of the communication system. 

In addition to performance, other criterions have to be considered when selecting a suitable virtualization method. These are the configuration complexity for setting up the respective system (configuration), which network interfaces are required (networking), whether automated deployment of multiple instances is possible (scalability), and the system's vulnerability (security level), where a threat model for attacker types A, B, C, and D was introduced in Sec. \ref{sec:Industrial Automation}. Therefore, Tab. \ref{tab1} sums up the key aspects for each network configuration.

\begin{table*}[htbp]
\caption{Comparison of the different network configurations}
\begin{center}
\begin{tabular*}{\textwidth}{|p{0.08\textwidth}||p{0.11\textwidth}|p{0.1\textwidth}|p{0.03\textwidth}|p{0.03\textwidth}|p{0.1\textwidth}|p{0.08\textwidth}|p{0.08\textwidth}|p{0.07\textwidth}|p{0.06\textwidth}|}
\cline{1-10}
\textbf{Network}& \multicolumn{2}{|c|}{\textbf{\textit{Config. of Host / Docker}}} & \multicolumn{2}{|c|}{\textbf{\textit{Networking}}} & \textbf{\textit{Scalability}} & \multicolumn{2}{|c|}{\textbf{\textit{Security Level}}}& \multicolumn{2}{|c|}{\textbf{\textit{Performance (RTT)}}}\\
\textbf{Config.} &  \textit{Host} & \textit{Docker} & \textit{L2} & \textit{L3} & & Guest/Host isolation & Network \linebreak isolation & \textit{Host-only} & \textit{Inter-Host} \\
\cline{1-10} 
\textit{Bare-Metal} & Not required & --- & Yes & Yes & No & --- & No & 51µs & 522µs\\
\cline{1-10} 
\textit{VM} & Config. of a Linux bridge per host & --- & Yes & Yes & Yes & Yes & No & 536µs & 992µs \\
\cline{1-10} 
\textit{Host} & Not required  & Not required & Yes & Yes &  No &  No & No & 53µs & 522µs \\
\cline{1-10} 
\textit{Bridge} & Config. of a Linux bridge per host & Config. of Docker network & Yes & Yes & Yes, but additional config. per Host  & No & Yes & 76µs & 600µs\\
\cline{1-10} 
\textit{Macvlan} & Only for Host-to-Container comm. & Config. of Docker network  & Yes & Yes & Yes, but additional config. per Host & Yes & No & 54µs & 520µs \\
\cline{1-10} 
\textit{Ipvlan (L2)} & Not required  & Config. of Docker network & No & Yes & Yes, but additional config. per Host & Yes & No & 55µs & 520µs  \\
\cline{1-10} 
\textit{Ipvlan (L3)} & Not required  & Config. of Docker network & No & Yes & Yes, but additional config. per Host & Yes & No & 58µs & 539µs  \\
\cline{1-10} 
\textit{Overlay} & Not required  & Not required & No & Yes & Yes & Yes & Yes & 71µs & 656µs\\
\cline{1-10} 
\end{tabular*}
\label{tab1}
\end{center}
\end{table*}

First the bare-metal should be evaluated. Here the performance is fine and the applications can directly access the physical network adapter of the system. Furthermore, this system allows \gls{l2} communication that is required by applications that belong to \gls{rt} class C. 
However, automated provisioning and configuration is not possible and there is no isolation, so security concerns are an important factor when using the infrastructure without virtualization.
In terms of network performance and isolation, \glspl{vm} behave in the opposite way compared to bare-metal. Performance is significantly worse compared to all other systems analyzed, but the security level is higher due to isolation when using a fully virtualized system. 
Therefore, \glspl{vm} can only be used if performance is not the primary focus.

Next, the different Docker network configurations should be discussed. In host mode the containers network can be accessed directly over the host’s adapter, exposing it to everything connected to the host 
and is therefore comparable to bare-metal. In terms of performance there is a small overhead given by the virtualization, but there is no scalability, since this network driver cannot be used for a Swarm service. In addition, 
attackers D and C can connect to the application if it is not authenticated and attackers A and B can monitor any traffic going over the host adapter, as host mode effectively lifts the networking isolation Docker offers by default. From a security perspective, this mode should only be used when necessary and with trusted services running on the container.

A Bridge network where the container connects to the host network over a virtual bridge is enabled by default. Nonetheless the official Docker documentation discourages from using default bridges, arguing that every unrelated container without a specified network communicates over the same bridge and may perform ARP spoofing or \gls{mac} flooding attacks. 
Assuming an isolated user-defined bridge is used, attacker B can only communicate to the host and over the specified ports, attacker A has full access to the bridge network. Attacker C and D only have access to the bridge network over ports forwarded by the host interface. User-defined bridges also provide a Swarm mode.

As macvlan makes the container appear as a physical device with an own \gls{mac} address, it is well suited for industrial communication. While a firewall can still restrict access by attacker D, attacker C can communicate with the device because the macvlan must be on the same subnet as the host interface. The traffic between the container and the underlying hosts is filtered by kernel modules to provide isolation, therefore access for attacker type A is restricted. Attackers of type B are isolated from the host but can access other containers macvlan network. If access by the host should not be allowed, configuration of the host is not required. For the use in a Docker  Swarm cluster, a configuration on Docker level is also required for this network driver once per host. 

Ipvlan L2 mode is functionally equivalent to the macvlan bridge mode, with the difference that all containers running on the same host are assigned the same \gls{mac} address. This may be useful when it is not possible to run the interface in promiscuous mode, which would be required for macvlan. In addition, 
certain network switches have a security feature where a physical port only transmits packets from one known \gls{mac} address to mitigate \gls{arp} spoofing attacks. In this case ipvlan is preferred over macvlan as all containers have the same \gls{mac} address and can share the secure port. Also, for whitelisting base MAC filters ipvlan may be preferred. Ipvlan L3 mode different subnets and networks can communicate when the parent interface is the same. Like in macvlan, direct communication with parent interface is not possible. In ipvlan L3 mode interfaces do not allow broadcast or multicast traffic, these are handled by the host. Routing, in contrary to L2 mode, is done by the parent interface according to the namespace, therefore L3 mode is better suited for non-trusted environments \cite{bandewar}. If an application requires \gls{l2} communication, and the security policies allow the assignment of a physical \gls{mac} address for each container, the macvlan network driver should be selected.

Overlay networks are the default network driver when deploying a Docker Swarm service and allow multiple Docker hosts to connect to a virtual network on top of an existing network. Therefore, the Docker daemons of each host are configured automatically 
This avoids misconfiguration, but causes an overhead that results in a loss of performance compared to the host, macvlan and ipvlan network drivers. By default, the traffic is not encrypted and can be seen by attacker type A, B, and C. Attacker D could also see unencrypted overlay traffic if two Docker daemons are connected over the internet and passes attacker D. Encryption can be easily enabled and would restrict attacker C and D from viewing the data. 

\subsection{Summary}
The investigations have shown that the use of containers leads to a better performance compared to \glspl{vm} in addition to the increased efficiency. Furthermore, container-based applications perform only slightly worse than bare-metal applications and can be used for industrial applications of each of the proposed \gls{rt} classes. A look at the computation performance shows that a Linux \gls{rt} kernel in combination with Docker can reliably run the cyclic execution of applications in intervals of \mbox{1µs - 1ms} with a jitter of  \flq 15µs. Comparing this with the \gls{rt} kernel without 
using Docker, the maximum jitter is about 1µs lower, which is almost negligible. From a security perspective, traffic encryption is recommended, but the non-negligible performance penalty does not allow network encryption for time-critical applications, then security must be achieved through a high degree of traffic isolation and other security measures. Looking deeper into the networking of each technology, we concluded that the macvlan network driver is best suited for applications that have high latency requirements and require \gls{l2} and \gls{l3} communications, while the overlay driver should be used when a higher level of security needs to be achieved and the application can handle \glspl{rtt} that are 20\% higher than those of other configurations.


\section{Conclusion}%
\label{sec:Conclusion}
In this paper, we have examined the possibilities of virtualization in the industrial landscape. Performance comparisons for bare-metal applications, \glspl{vm} and containers were conducted and evaluated for industrial needs.  Furthermore, the available network drivers of Docker containers were listed and compared based on configuration effort, accessibility, scalability, security level, and performance. Again, the performance for \glspl{vm} and applications without virtualization technology was compared. The analyses indicate that the flexibility gained by virtualization can be achieved for industrial applications without violating the stringent \gls{rt} requirements in the industrial landscape, if Docker containers with a suitable configuration are used.


\printbibliography%

@online{b1,
	KEYWORDS		= {primary_literature},
	AUTHOR			= {H. Kagermann AND W.-D. Lukas  AND W. Wahlster},
	TITLE			= {Industrie 4.0: Mit dem Internet der Dinge auf dem Weg zur 4. industriellen Revolution},
	YEAR			= {2011},
    month			= {April},
    %URL			= {https://www.vdi-nachrichten.com/Technik-Gesellschaft/Industrie-40-Mit-Internet-Dinge-Weg-4-industriellen-Revolution},
 %   NOTE			= {Last accessed: 2018-03-13},
}

@InProceedings{7095802,
  author    = {W. Felter and A. Ferreira and R. Rajamony and J. Rubio},
  title     = {An updated performance comparison of virtual machines and Linux containers},
  booktitle = {2015 IEEE International Symposium on Performance Analysis of Systems and Software (ISPASS)},
  year      = {2015},
  pages     = {171-172},
  month     = {March},
  doi       = {10.1109/ISPASS.2015.7095802},
}

@INPROCEEDINGS{7883994, 
author={M. Wollschlaeger and T. Sauter and J. Jasperneite}, 
journal={IEEE Industrial Electronics Magazine}, 
title={The Future of Industrial Communication: Automation Networks in the Era of the Internet of Things and Industry 4.0}, 
year={2017}, 
volume={11}, 
number={1}, 
pages={17-27}, 
doi={10.1109/MIE.2017.2649104}, 
ISSN={1932-4529}, 
month={March},}

@INPROCEEDINGS{6837587, 
author={O. Givehchi and J. Imtiaz and H. Trsek and J. Jasperneite}, 
booktitle={2014 10th IEEE Workshop on Factory Communication Systems (WFCS 2014)}, 
title={Control-as-a-service from the cloud: A case study for using virtualized PLCs}, 
year={2014}, 
volume={}, 
number={}, 
pages={1-4}, 
doi={10.1109/WFCS.2014.6837587}, 
ISSN={}, 
month={May},}

@INPROCEEDINGS{8502649, 
author={M. Gundall and J. Schneider and H. D. Schotten and M. Aleksy and D. Schulz and N. Franchi and N. Schwarzenberg and C. Markwart and R. Halfmann and P. Rost and D. Wübben and A. Neumann and M. Düngen and T. Neugebauer and R. Blunk and M. Kus and J. Grießbach}, 
booktitle={2018 IEEE 23rd International Conference on Emerging Technologies and Factory Automation (ETFA)}, 
title={5G as Enabler for Industrie 4.0 Use Cases: Challenges and Concepts}, 
year={2018}, 
volume={1}, 
number={}, 
pages={1401-1408}, 
doi={10.1109/ETFA.2018.8502649}, 
ISSN={1946-0759}, 
month={Sep.},}

@ARTICLE{6246692,
author={P. Gaj and J. Jasperneite and M. Felser},
journal={IEEE Transactions on Industrial Informatics},
title={Computer Communication Within Industrial Distributed Environment—a Survey},
year={2013},
volume={9},
number={1},
pages={182-189},
doi={10.1109/TII.2012.2209668},
ISSN={1551-3203},
month={Feb},}

@INPROCEEDINGS{ldd,
	AUTHOR			= {K. Pietrusewicz and L. Urbanski},
	TITLE			= {Control programming software strategies for industrial systems},
	YEAR			= {2011},
    month			= {},
    %URL			= {},
    %NOTE			= {Last accessed:},
}

@INPROCEEDINGS{microvirtualization,
	AUTHOR			= {Bromium},
	TITLE			= {Whitepaper: Micro-virtualization vs Software Sandboxing},
	YEAR			= {2013},
    month			= {},
    %URL			= {},
    %NOTE			= {Last accessed:},
}

@ARTICLE{8401919, 
author={E. {Sisinni} and A. {Saifullah} and S. {Han} and U. {Jennehag} and M. {Gidlund}}, 
journal={IEEE Transactions on Industrial Informatics}, 
title={Industrial Internet of Things: Challenges, Opportunities, and Directions}, 
year={2018}, 
volume={14}, 
number={11}, 
pages={4724-4734}, 
doi={10.1109/TII.2018.2852491}, 
ISSN={1551-3203}, 
month={Nov},}

@INPROCEEDINGS{8254628, 
author={S. Tayade and P. Rost and A. Maeder and H. D. Schotten}, 
booktitle={GLOBECOM 2017 - 2017 IEEE Global Communications Conference}, 
title={Device-Centric Energy Optimization for Edge Cloud Offloading}, 
year={2017}, 
volume={}, 
number={}, 
pages={1-7}, 
doi={10.1109/GLOCOM.2017.8254628}, 
ISSN={}, 
month={Dec},}

@ARTICLE{8399557,
author={L. {Yin} and J. {Luo} and H. {Luo}},
journal={IEEE Transactions on Industrial Informatics},
title={Tasks Scheduling and Resource Allocation in Fog Computing Based on Containers for Smart Manufacturing},
year={2018},
volume={14},
number={10},
pages={4712-4721},
doi={10.1109/TII.2018.2851241},
ISSN={1941-0050},
month={Oct},}

@article{Mao2017MobileEC,
  title={Mobile Edge Computing: Survey and Research Outlook},
  author={Yuyi Mao and Changsheng You and Jun Zhang and Kaibin Huang and Khaled Ben Letaief},
  journal={ArXiv},
  year={2017},
  volume={abs/1701.01090}
}

@article{10.1145/2723872.2723882,
author = {Boettiger, Carl},
title = {An Introduction to Docker for Reproducible Research},
year = {2015},
issue_date = {January 2015},
publisher = {Association for Computing Machinery},
address = {New York, NY, USA},
volume = {49},
number = {1},
issn = {0163-5980},
url = {https://doi.org/10.1145/2723872.2723882},
doi = {10.1145/2723872.2723882},
journal = {SIGOPS Oper. Syst. Rev.},
month = jan,
pages = {71–79},
numpages = {9}
}

@inproceedings{10.1145/2851613.2851737,
author = {Moga, Alexandru and Sivanthi, Thanikesavan and Franke, Carsten},
title = {OS-Level Virtualization for Industrial Automation Systems: Are We There Yet?},
year = {2016},
isbn = {9781450337397},
publisher = {Association for Computing Machinery},
address = {New York, NY, USA},
url = {https://doi.org/10.1145/2851613.2851737},
doi = {10.1145/2851613.2851737},
booktitle = {Proceedings of the 31st Annual ACM Symposium on Applied Computing},
pages = {1838–1843},
numpages = {6},
location = {Pisa, Italy},
series = {SAC ’16}
}

@article{Combe.2016,
 author = {Combe, Theo and Martin, Antony and {Di Pietro}, Roberto},
 year = {2016},
 title = {To Docker or Not to Docker: A Security Perspective},
 pages = {54--62},
 volume = {3},
 number = {5},
 journal = {IEEE Cloud Computing},
 doi = {10.1109/MCC.2016.100},
}

@article{uhlig2005intel,
  title={Intel virtualization technology},
  author={Uhlig, Rich and Neiger, Gil and Rodgers, Dion and Santoni, Amy L and Martins, Fernando CM and Anderson, Andrew V and Bennett, Steven M and Kagi, Alain and Leung, Felix H and Smith, Larry},
  journal={Computer},
  volume={38},
  number={5},
  pages={48--56},
  year={2005},
  publisher={IEEE}
  }

@INPROCEEDINGS{7164727, author={A. M. {Joy}}, booktitle={2015 International Conference on Advances in Computer Engineering and Applications}, title={Performance comparison between Linux containers and virtual machines}, year={2015}, volume={}, number={}, pages={342-346},}

@INPROCEEDINGS{1137388, author={L. {Abeni} and A. {Goel} and C. {Krasic} and J. {Snow} and J. {Walpole}}, booktitle={Proceedings. Eighth IEEE Real-Time and Embedded Technology and Applications Symposium}, title={A measurement-based analysis of the real-time performance of linux}, year={2002}, volume={}, number={}, pages={133-142},}

@article{yodaiken1997real,
  title={Real-time linux applications and design},
  author={Yodaiken, Victor and Barabanov, Michael}
}

@inproceedings{dietrich2005evolution,
  title={The evolution of real-time linux},
  author={Dietrich, Sven-Thorsten and Walker, Daniel},
  booktitle={7th RTL Workshop},
  year={2005},
  organization={Citeseer}
}

@article{goldschmidt2018container,
  title={Container-based architecture for flexible industrial control applications},
  author={Goldschmidt, Thomas and Hauck-Stattelmann, Stefan and Malakuti, Somayeh and Gr{\"u}ner, Sten},
  journal={Journal of Systems Architecture},
  volume={84},
  pages={28--36},
  year={2018},
  publisher={Elsevier}
}

@inproceedings{opcuart,
author = {Pfrommer, Julius and Ebner, Andreas and Ravikumar, Siddharth and Karunakaran, Bhagath},
year = {2018},
month = {07},
pages = {},
title = {Open Source OPC UA PubSub over TSN for Realtime Industrial Communication},
doi = {10.1109/ETFA.2018.8502479}
}

@INPROCEEDINGS{8731776,
author={C. {Mannweiler} and B. {Gajic} and P. {Rost} and R. S. {Ganesan} and C. {Markwart} and R. {Halfmann} and J. {Gebert} and A. {Wich}},
booktitle={Mobile Communication - Technologies and Applications; 24. ITG-Symposium},
title={Reliable and Deterministic Mobile Communications for Industry 4.0: Key Challenges and Solutions for the Integration of the 3GPP 5G System with IEEE},
year={2019},
volume={},
number={},
pages={1-6},
doi={},
ISSN={},
month={5},}

@article{marmol2015networking,
  title={Networking in containers and container clusters},
  author={Marmol, Victor and Jnagal, Rohit and Hockin, Tim},
  journal={Proceedings of netdev 0.1},
  year={2015},
}

@article{s7300,
  title={SIMATIC S7-300C: Die kompakten Controller für die Fertigungsautomatisierung bieten vielfältige Innovationen und integrierte Funktionen},
  author={Siemens AG},
  year={2011},
  %publisher={Siemens AG}
  URL = {https://c4b.gss.siemens.com/resources/images/articles/e20001-a760-p210.pdf},
}

@article{s71500,
    title={Delivery Release for the new SIMATIC S7-1500 Controllers},
    author={Siemens AG},
    URL={https://support.industry.siemens.com/cs/document/67856446/},
    year={2013},
    month={3},
}

@article{Zhu.2011,  author={B. {Zhu} and A. {Joseph} and S. {Sastry}},  booktitle={2011 International Conference on Internet of Things and 4th International Conference on Cyber, Physical and Social Computing},  title={A Taxonomy of Cyber Attacks on SCADA Systems},   year={2011},  volume={},  number={},  pages={380-388},}

@inproceedings{bandewar,
author = {Bandewar, Mahesh and Dumazet, Eric},
year = {2015},
month = {02},
pages = {},
booktitle={Proceedings of netdev 0.1}, 
title = {IPVLAN - The beginning},
%doi = {10.1109/ETFA.2018.8502479}
}
\nl
\TempDisplayPreparation
\end{document}